\documentclass[aps,
               prl,
               showpacs,
               amssymb,
               nofootinbib,
               superscriptaddress,
               twocolumn,
               longbibliography,10pt]{revtex4-2}

\usepackage[ansinew]{inputenc}
\usepackage{bbold}
\usepackage{bbm}
\usepackage{bm}
\usepackage{amsbsy}
\usepackage{amsthm}
\usepackage{amssymb}
\usepackage{amsfonts}
\usepackage{algorithm}
\usepackage{algpseudocode}
\usepackage{amsmath, mathtools}
\usepackage{dsfont}
\usepackage{graphicx}
\usepackage{epsfig}
\usepackage{epstopdf}
\usepackage{mathrsfs} 

\usepackage[english]{babel}
\babeladjust{autoload.bcp47 = on}
\usepackage{array}

\usepackage[dvipsnames]{xcolor}
\usepackage[colorlinks]{hyperref}
\makeatletter
\newcommand\org@hypertarget{}
\let\org@hypertarget\hypertarget
\renewcommand\hypertarget[2]{%
  \Hy@raisedlink{\org@hypertarget{#1}{}}#2%
  }
\makeatother
\usepackage[figure,table]{hypcap}

\usepackage{MnSymbol}
\usepackage{enumerate}
\usepackage{float}
\usepackage{comment}
\usepackage{braket}
\usepackage{subfigure}
\usepackage{orcidlink}
\usepackage{tcolorbox}
\makeatletter
\newcommand{\providetcbcountername}[1]{%
  \@ifundefined{c@tcb@cnt@#1}{%
    --undefined--%
  }{%
    tcb@cnt@#1%
  }
}
\newcommand{\ketbra}[2]{\ensuremath{|{#1}\rangle\!\langle{#2}|}}

\newcommand{\settcbcounter}[2]{%
  \@ifundefined{c@tcb@cnt@#1}{%
    \GenericError{Error}{counter name #1 is no tcb counter }{}{}%
  }{%
    \setcounter{tcb@cnt@#1}{#2}%
   }%
}%

\usepackage{subfigure}
\usepackage{makecell}

\hypersetup{
	bookmarksnumbered,
	pdfstartview={FitH},
	citecolor={darkgreen},
	linkcolor={darkred},
	urlcolor={darkblue},
	pdfpagemode={UseOutlines}}
\definecolor{darkgreen}{RGB}{50,190,50}
\definecolor{darkblue}{RGB}{0,0,190}
\definecolor{darkred}{RGB}{238,0,0}
\definecolor{mycolor}{rgb}{0.122, 0.435, 0.698}
\usepackage{soul}

\usepackage{tabularx}

\newcommand{\subtiny}[3]{\ensuremath{_{\hspace{#1 pt}\protect\raisebox{#2 pt}{\tiny{$ #3$}}}}}

\newcommand{\tr}{\textnormal{Tr}}

\renewcommand{\thesection}{\Roman{section}}
\renewcommand{\thesubsection}{\Roman{section}.\Alph{subsection}}
\renewcommand{\thesubsubsection}{\Roman{section}.\Alph{subsection}.\arabic{subsubsection}}
\makeatletter
\renewcommand{\p@subsection}{}
\renewcommand{\p@subsubsection}{}
\makeatother

\begin{document}
\title{Experimental verification of multi-copy activation of genuine multipartite entanglement}

\author{Robert St\'{a}rek\,\orcidlink{0000-0002-5396-6293}}
\email{starek@optics.upol.cz}
\affiliation{Department of Optics, Palack\'y University, 17. listopadu 1192/12, 779~00 Olomouc, Czech Republic}
\author{Tim Gollerthan\,\orcidlink{0009-0005-5658-6017}}
\affiliation{Universit\"at Innsbruck, Institut f\"ur Experimentalphysik, 6020 Innsbruck, Austria}
\author{Olga Leskovjanov{\'a}\,\orcidlink{0000-0002-3638-3627}}
\affiliation{Department of Optics, Palack\'y University, 17. listopadu 1192/12, 779~00 Olomouc, Czech Republic}
\author{Michael Meth\,\orcidlink{0000-0002-5446-3962}}
\affiliation{Universit\"at Innsbruck, Institut f\"ur Experimentalphysik, 6020 Innsbruck, Austria}
\author{Peter Tirler\,\orcidlink{0009-0000-2123-666X}}
\affiliation{Universit\"at Innsbruck, Institut f\"ur Experimentalphysik, 6020 Innsbruck, Austria}
\author{Nicolai Friis\,\orcidlink{0000-0003-1950-8640}}
\affiliation{Technische Universit{\"a}t Wien, Atominstitut \& Vienna Center for Quantum Science and Technology (VCQ),  Stadionallee 2, 1020 Vienna, Austria}
\author{Martin Ringbauer\,\orcidlink{0000-0001-5055-6240}}
\affiliation{Universit\"at Innsbruck, Institut f\"ur Experimentalphysik, 6020 Innsbruck, Austria}
\author{Ladislav Mi\v{s}ta, Jr.\,\orcidlink{0000-0002-1979-7617}}
\email{mista@optics.upol.cz}
\affiliation{Department of Optics, Palack\'y University, 17. listopadu 1192/12, 779~00 Olomouc, Czech Republic}

\date{\today}

\begin{abstract} 
A central concept in quantum information processing is genuine multipartite entanglement (GME), a type of correlation beyond biseparability, that is, correlations that cannot be explained by statistical mixtures of partially separable states. GME is relevant for characterizing and benchmarking complex quantum systems, and it is an important resource for applications such as quantum communication. Remarkably, it has been found that GME can be activated from multiple copies of biseparable quantum states, which do not possess GME individually. Here, we experimentally demonstrate unambiguous evidence of such GME activation from two copies of a biseparable three-qubit state in a trapped-ion quantum processor. These results not only challenge notions of quantum resources but also highlight the potential of using multiple copies of quantum states to achieve tasks beyond the capabilities of the individual copies.
\end{abstract}

\date{\today}

\maketitle


A key goal in the development of quantum-communication technology is to establish large-scale quantum networks~\cite{Kimble2008,Duer2017,Simon2017,WehnerElkoussHanson2018,Cacciapuoti2019}. 
Central questions in this endeavor pertain to understanding what kind of quantum states can be established by specific networks~\cite{NavascuesWolfeRossetPozasKerstjens2020,NavascuesWolfe2020,KraftDesignolleRitzBrunnerGuehneHuber2021,WolfePozasKerstjensGrinbergRossetAcinNavascues2021}, which resources are required to do so, and how the successful generation can be efficiently verified~\cite{KraftSpeeYuGuehne2021,HansenneXuKraftGuehne2022,LiDaiMunozAriasReuerHuberFriis2025}. 
A particular focus of these efforts (both on the side of theory, see, e.g.,~\cite{ContrerasTejadaPalazuelosDeVicente2021,ContrerasTejadaPalazuelosDeVicente2022,MorelliSauerweinSkotiniotisFriis2022} and experiments, for instance~\cite{BesseEtAl2020,PompiliEtAl2021,Ruskuc2025,shi2025,CanteriBateMishraFriisKrutyanskiyLanyon2025}) is the generation of \emph{genuinely multipartite entangled} states, needed to harness the full potential of quantum networks. 
Such states are not just fully inseparable in the sense that they are entangled across all bipartitions, they also cannot be decomposed into statistical mixtures of states that are separable with respect to different partitions, whereas all states that admit such decompositions are called biseparable. For an introduction, we refer to~\cite[Chapter~18]{BertlmannFriis2023}.  
Multipartite entanglement is considered to be an important resource for tasks in quantum metrology~\cite{Toth2012}, quantum computing (e.g., for measurement-based quantum computation~\cite{RaussendorfBriegel2001, BriegelRaussendorf2001} and quantum error correction~\cite{Scott2004}), and quantum communication (e.g., for quantum key distribution~\cite{EppingKampermannMacchiavelloBruss2017,PivoluskaHuberMalik2018}, conference key agreement~\cite{RibeiroMurtaWehner2018}, or communication problems in networks~\cite{BaeumlAzuma2017}), and there are some applications for which genuine multipartite entanglement (GME) specifically is crucial~\cite{EppingKampermannMacchiavelloBruss2017, YamasakiPirkerMuraoDuerKraus2018}.

Remarkably, it has been shown that considering more than one copy of a state drastically changes the distinction between full inseparability and GME~\cite{HuberPlesch2011, YamasakiMorelliMiethlingerBavarescoFriisHuber2022}: Where one copy of a state may be biseparable, two or more copies can be GME as long as the single-copy state is fully inseparable (i.e., entangled with respect to all bipartitions)---a phenomenon dubbed \emph{activation of GME}. Moreover, it was shown that every fully inseparable biseparable state can be activated for some number of copies~\cite{PalazuelosDeVicente2022}, even in infinite dimensions~\cite{BaksovaLeskovjanovaMistaAgudeloFriis2025}. 

A pressing question that follows on the heel of these observations is: \emph{How difficult is it to harness the activation of GME?} 
Theoretical work in this direction~\cite{WeinbrennerEtAl2024} has already demonstrated that there are some fully inseparable biseparable states whose activated GME cannot be projected back to the single-copy level. 
In addition, some cases might require prohibitively many copies for activation, and some states with activated GME might require joint local operations on multiple copies that are difficult to realize in practice in order to verify or use the activated multipartite correlations. 

Here, we make crucial steps towards bringing the utilization of GME activation closer to practical reality by unambiguously demonstrating its core principle: We experimentally prepare two copies of a biseparable three-qubit state on two groups of three trapped ions, and show that the two-copy state is genuinely multipartite entangled. Whereas previous work in this direction~\cite{ChenEtAl2024} only checked necessary (but not sufficient) conditions for biseparability of the individual copies, state preparation in our experiment solely employs operations that cannot produce genuinely multipartite entangled three-qubit states or bipartite entanglement between the two single-copy instances of the three-qubit states. In addition, explicit biseparable decompositions for the initial three-qubit states are determined by a numerical algorithm~\cite{Barreiro2010, Hofmann2014}. We confirm the activation of GME by employing a suitable GME witness for the two-copy state.
Our results thus provide clear evidence of two-copy GME activation. This marks a significant step in the exploration of quantum resources that can be harnessed by jointly but locally accessing multiple copies of distributed quantum states in the laboratory.\\ 


\noindent\emph{Theory}.\ 
In order to experimentally demonstrate multi-copy GME activation, we consider a state $\rho\subtiny{0}{0}{A\hspace*{-0.5pt}BC}$ of three qubits that is biseparable, i.e., can be written as a statistical mixture of states $\rho\subtiny{0}{0}{A|BC}$, $\rho\subtiny{0}{0}{A\hspace*{-0.5pt}B|C}$, and $\rho\subtiny{0}{0}{B|AC}$ that are separable with respect to the bipartitions $A|BC$, $AB|C$ and $B|AC$, respectively, but which is fully inseparable, that is, $\rho\subtiny{0}{0}{A\hspace*{-0.5pt}BC}$ cannot be written as a statistical mixture of terms that are all separable with respect to any fixed bipartition. 
Yet, $\rho\subtiny{0}{0}{A\hspace*{-0.5pt}BC}$ is two-copy activatable: Two copies $\rho\subtiny{0}{0}{A_{1}\hspace*{-0.5pt}B_{1}C_{1}}\otimes\rho\subtiny{0}{0}{A_{2}B_{2}C_{2}}$ of this state are GME, i.e., the joint two-copy state is not biseparable with respect to the partition $A_{1}A_{2}|B_{1}B_{2}|C_{1}C_{2}$. 
In addition, we are interested in a state that is sufficiently robust with respect to these properties: Small perturbations should not change the biseparability and full inseparability of the single copy or the GME of two copies. 
We are also interested in an implementation that is closest to the original spirit of activation, namely that each copy is prepared directly as a convex mixture of product states without using any operation that could potentially generate GME. However, the potentially large number of states in such mixtures, combined with the need to prepare multiple copies, might lead to significant overhead in terms of the number of required reconfigurations of the experimental setup. To keep this number at a level achievable with current technology, we construct an activatable state in which the number of components is sufficiently small.

By combining analytical and numerical calculations, we arrived at a suitable candidate for the desired state in the form of the balanced mixture
\begin{equation}
    \tilde{\rho}\subtiny{0}{0}{A\hspace*{-0.5pt}BC} =\tfrac{1}{8}\sum\limits_{i=0}^{7}|a_i\rangle\!\langle a_i|
    \label{eq:rho}
\end{equation}
of only eight three-qubit states $\ket{a_{i}}$, $i=0,1,\ldots,7$. The first four of these have the form 
\begin{subequations}
    \label{a}
\begin{align}
|a_{0,1}\rangle &= |\pm\rangle\subtiny{0}{0}{A} \otimes |\Phi^{\pm}\rangle\subtiny{0}{0}{BC}, \\[1mm]
|a_{2,3}\rangle &= \bigl(\sqrt{Z}\otimes\sqrt{Z}\otimes Z \bigr)|\pm\rangle\subtiny{0}{0}{A} \otimes |\Phi^{\pm}\rangle\subtiny{0}{0}{BC},
\end{align}
\end{subequations}
while the remaining four, $|a_{4,\dots,7}\rangle$, arise from them by swapping qubits $A$ and $B$. Here, the subscripts $0$ and $2$, and $1$ and $3$ on the left-hand side refer to the signs $+$ and $-$ on the right-hand side, $\ket{\Phi^\pm}=(\ket{00}\pm\ket{11})/\sqrt{2}$, $\ket{\pm}=(\ket{0}\pm\ket{1})/\sqrt{2}$, $Z$ is the Pauli-$z$ matrix, and $\sqrt{Z}=\operatorname{diag}(1,i)$.

The state defined in Eq.~\eqref{eq:rho} is clearly a convex mixture of separable states with respect to the partitions $A|BC$ and $B|AC$, and is thus biseparable by construction as required. In addition, two copies of the state (\ref{eq:rho}), $\tilde{\rho}\subtiny{0}{0}{A_{1}\hspace*{-0.5pt}B_{1}C_{1}}\otimes\tilde{\rho}\subtiny{0}{0}{A_{2}B_{2}C_{2}}$, are GME across the partition $A_{1}A_{2}|B_{1}B_{2}|C_{1}C_{2}$, as shown below, and the state is thus also two-copy GME activatable. This is somewhat surprising when we realize that we only need a \emph{single} two-qubit CNOT operation to prepare each constituent of the state (\ref{eq:rho}), which obviously cannot generate GME. 
Experimental activation of GME based on the state (\ref{eq:rho}) would thus represent a practically ideal demonstration of this counterintuitive effect of obtaining ``something from nothing.'' But before we move on to that, let us first prove GME in two copies of the state (\ref{eq:rho}).

The GME can be shown by finding a witness with respect to the partition $A_{1}A_{2}|B_{1}B_{2}|C_{1}C_{2}$ for its two-copy state which we for convenience rearrange as $\tilde{\rho}\subtiny{0}{0}{A_{1}A_{2}B_{1}B_{2}C_{1}C_{2}}$. In general, a GME witness is a Hermitian operator $W$ for which $\tr[W\rho^{\rm bisep}]\geq 0$ for all biseparable states $\rho^{\rm bisep}$ and $\tr[W\rho]<0$ for at least one GME state $\rho$. The GME of a number of states, including two copies of the state $\tilde{\rho}\subtiny{0}{0}{A\hspace*{-0.5pt}BC}$ [Eq.~\eqref{eq:rho}], can be detected using so-called fully decomposable witnesses, which can be written as~\cite{JungnitschMoroderGuehne2011a} 
\begin{equation}
    W = P_M + Q_M^{T_M},
\end{equation}
for every subset $M$ of all systems.
Here, $P_M$ and $Q_M$ are positive semi-definite matrices and the superscript $T_M$ denotes the partial transposition with respect to the part $M$ of the whole system~\cite{Peres1996, Horodecki1997}. The important upside of a fully decomposable witness is that it can be found relatively simply by solving a semi-definite program (SDP)~\cite{JungnitschMoroderGuehne2011a}. In our case, the SDP is 
\begin{equation}\label{prog:sdp}
\begin{aligned}
& \underset{W,P_{k}}{\text{minimize}}
& & \tr[W\tilde{\rho}\subtiny{0}{0}{A_{1}A_{2}B_{1}B_{2}C_{1}C_{2}}] \\
& \text{subject to}
& & \tr [W] = 1,\\
& & & P_{k}\geq0,\\
& & & Q_k = \left(W - P_k\right)^{T_k}\geq 0,\\
& & & \textrm{for} \quad  k = \{A_1 A_2, B_1 B_2, C_1 C_2 \}.
\end{aligned}
\end{equation}

We solve this SDP numerically and obtain $\tr[W \tilde{\rho}\subtiny{0}{0}{A_{1}A_{2}B_{1}B_{2}C_{1}C_{2}} ]=-1.042\cdot10^{-2}$, verifying that the witness faithfully detects GME of two copies of the state $\tilde{\rho}\subtiny{0}{0}{A\hspace*{-0.5pt}BC}$ as desired. The witness $W$ has non-zero matrix elements only on the main diagonal and anti-diagonal, with elements $1/12$ that are listed explicitly together with the matrices $P_{M}$ and $Q_{M}$ in Appendix~\ref{appendix:1}. Additionally, $W$ can be decomposed into a sum of 32 six-qubit Pauli strings $M_{k}$,
\begin{equation}
    W = \sum\limits_{k=0}^{31} m_k M_k\,,
    \label{eq:decomposedwitness}
\end{equation}
with weights $m_{k}$ (see Appendix~\ref{appendix:1} for the list of these products and their weights). Half of the matrices $M_{k}$ correspond to computational basis measurements of various subsets and can thus be measured at once. Hence, this decomposition is experimentally very convenient and allows for a direct witness measurement using only 17 distinct measurement settings. This is much less expensive than the $3^6=729$ settings required for Pauli state tomography.

\begin{figure*}[ht]
    \centering
    \includegraphics[scale=1]{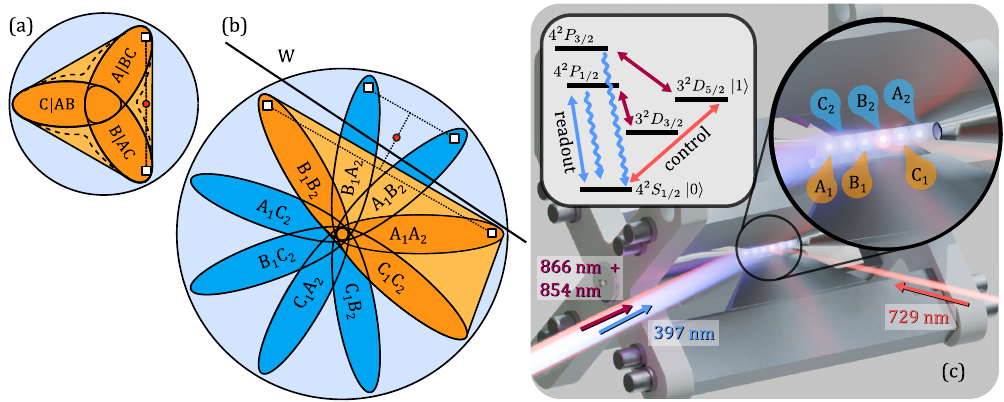}
    \caption{\textbf{(a)} Illustration of the robust biseparable two-copy GME-activatable state $\rho\subtiny{0}{0}{A\hspace*{-0.5pt}BC}$ [Eq.~\eqref{eq:ourrho}]. The dark orange oval regions represent the sets of separable states across bipartitions $A|BC, B|AC$, and $C|AB$. The light orange regions represent the 
    GME-activatable states. The union of the orange regions represents the set of biseparable states, while the light-blue region outside contains the GME states. The orange regions between the dashed lines and the borders between the sets of biseparable and GME states represent the set of biseparable two-copy GME-activatable states. The state $\rho\subtiny{0}{0}{A\hspace*{-0.5pt}BC}$ (red dot) is a balanced mixture (illustrated by the dotted line) of states separable across bipartitions $A|BC$ and $B|AC$ (white squares), respectively.\\
    \textbf{(b)} Diagrammatic representation of two copies of the state $\rho\subtiny{0}{0}{A\hspace*{-0.5pt}BC}$. The oval regions labeled by $jk$ with $j=A_{1}, B_{1}, C_{1}$ and $k=A_{2}, B_{2}, C_{2}$ denote sets of states for which the pair of qubits $j$ and $k$ is separable from the rest of the system. The convex hull of the three oval orange regions indicates the set of biseparable states with respect to the partition $A_{1}A_{2}|B_{1}B_{2}|C_{1}C_{2}$. 
    The two-copy state $\rho\subtiny{0}{0}{A_{1}A_{2}B_{1}B_{2}C_{1}C_{2}}$
     (red dot) is a balanced mixture (illustrated by the dotted lines) of four possible tensor products of states depicted by white squares in panel~(a), which belong to the sets $A_{1}A_{2}$, $B_{1}B_{2}$, $A_{1}B_{2}$, and $B_{1}A_{2}$ (white squares), respectively. GME with respect to the partition  $A_{1}A_{2}|B_{1}B_{2}|C_{1}C_{2}$ is detected by the witness $W$ (solid black line). \textbf{(c)} Illustration of a linear Paul trap and a $^{40}\mathrm{Ca}^{+}$ level diagram. A Paul trap consisting of four blade electrodes and two tip electrodes confines a linear chain of six $^{40}\mathrm{Ca}^{+}$ ions (white dots). The orange and blue labels illustrate the interleaved qubit assignment of the first and the second copy. We refer to the main text for details on the energy-level diagram.}
    \label{fig:fig1}
\end{figure*}

The state in Eq.~\eqref{eq:rho} lies close to the boundary of the set of biseparable states, and the biseparability condition is thus highly sensitive to experimental errors. This obstacle can be circumvented by admixing a small fraction $q=6\cdot 10^{-2}$ of colored noise in the form of $\bigl(|\tilde{a}_{8}\rangle\!\langle \tilde{a}_{8}|+|\tilde{a}_{9}\rangle\!\langle \tilde{a}_{9}|\bigr)/2$ with $|\tilde{a}_8\rangle= |001\rangle\subtiny{0}{0}{A\hspace*{-0.5pt}BC}$ and $|\tilde{a}_9\rangle= |110\rangle\subtiny{0}{0}{A\hspace*{-0.5pt}BC}$ to the state in Eq.~\eqref{eq:rho}. The exact value of $q=6\cdot10^{-2}$ was selected based on the preliminary analysis of the single-copy states in our experiment, as is explained in more detail in Appendix~\ref{appendix:2}. 
This gives 
\begin{equation}
\rho\subtiny{0}{0}{A\hspace*{-0.5pt}BC} = 
\tfrac{1-q}{8}\sum\limits_{i=0}^{7} |a_i\rangle\!\langle a_i|\,+\,\tfrac{q}{2}\bigl(|\tilde{a}_{8}\rangle\!\langle \tilde{a}_{8}|+|\tilde{a}_{9}\rangle\!\langle \tilde{a}_{9}|\bigr).
    \label{eq:ourrho}
\end{equation}
The obtained state is simple, manifestly biseparable and experimentally robust [see Fig.~\ref{fig:fig1}(a) for a pictorial representation]. For the GME witness $W$ one further expects $\langle W\rangle=\tr[W \rho\subtiny{0}{0}{A_{1}A_{2}B_{1}B_{2}C_{1}C_{2}}]=-0.887\cdot10^{-2}$, which certifies two-copy GME activatability of the state, as illustrated in Fig.~\ref{fig:fig1}(b). \\

\noindent\emph{Experimental GME activation}.\ We prepared two copies of the state in Eq.~\eqref{eq:ourrho} on a trapped-ion quantum processor~\cite{Ringbauer2022}. It employs a linear Paul trap, where six $^{40}\mathrm{Ca}^{+}$ ions -- three for each copy -- were confined [see Fig.~\ref{fig:fig1}(c)]. Qubits are encoded in the electronic states $|0\rangle=4^2S_{1/2}(m_{j}=-1/2)$ and $|1\rangle=3^2D_{5/2}(m_{j}=-1/2)$ and are coherently controlled via a narrowband laser driving an electric quadrupole transition at $729\,$nm. As illustrated in Fig.~\ref{fig:fig1}(c), the short-lived $4^2P_{1/2}$ state and the $4^2S_{1/2}$ state are coupled via a $397\,$nm laser, which allows for the effective implementation of both Doppler and polarization-gradient cooling. Ions spontaneously decaying to the $3^2D_{3/2}$ level are pumped back into the cooling cycle with a $866\,$nm laser. Furthermore, another laser at $854\,$nm permits population transfer from $3^2D_{5/2}$ to the short-lived $4^2P_{3/2}$ state, which decays back to $4^2S_{1/2}$. This closed cycle enables cooling to the motional ground state via resolved sideband cooling. 

State preparation control sequences include individually addressed qubit rotations around arbitrary axes in the equatorial plane of the Bloch sphere, virtual $Z$ gates, and entangling M{\o}lmer{\textendash}S{\o}rensen (MS) gates applied to arbitrary pairs of qubits \cite{SoerensenMoelmer1999}. The latter were used to generate the Bell states $|\Phi^{\pm}\rangle$ in Eq.~\eqref{a} from the ground state $|00\rangle$. The two copies of the three-qubit states in Eq.~\eqref{eq:ourrho} were prepared on ions 0, 2, and 4  as well as on ions 5, 3, and 1, respectively, with ions indexed according to their positioning in the trap. This ordering is chosen to reduce imperfections in the addressing of neighboring ions.
The $\sqrt{Z}$ gates of Eq.~\eqref{a} and swap gates were implemented virtually -- the former by adjusting the phase of consecutive pulses, the latter by relabelling the ions. Projective measurements are performed at the end of the gate sequence by driving the $4^2S_{1/2}$ to $4^2P_{1/2}$ transition with a $397\,$nm laser and collecting the fluorescence. Ions in the states $|0\rangle$ and $|1\rangle$ are discriminated by their respective bright or dark appearance with a readout error below $2\cdot10^{-3}$.

We performed tomographic characterization separately for each of the ten prepared three-qubit states, $|a_{i}\rangle$, $i=0,\ldots,7$, and $|\tilde{a}_{j}\rangle$, $j=8,9$, using Pauli tomography with 200 shots per measurement configuration and constituent state. The constituent state density matrices were then individually reconstructed via a maximum-likelihood estimation algorithm~\cite{Hradil2004}, and subsequently incoherently mixed as in Eq.~\eqref{eq:ourrho} with the mixing factor of $q=6\cdot 10^{-2}$. Uncertainties were calculated using 100 runs of Monte Carlo resampling. From these tomographies, we extract infidelities $1-F$ of the two single-copy states with the target state in Eq.~\eqref{eq:ourrho} of $(2.54 \pm 0.05)\cdot 10^{-2}$ and $(1.30 \pm 0.06)\cdot 10^{-2}$, respectively.

Next, we certified the biseparability of the single copies. For this purpose, we used the numerical algorithm of Ref.~\cite{Barreiro2010, Hofmann2014}, based on the sequential subtraction of product states from the investigated density matrix, which is described in detail in Appendix~\ref{appendix:2}. The algorithm effectively decomposes the original density matrix into a sum of product states and a small fully separable remainder, thereby proving its biseparability.

Due to the structure of state $\rho\subtiny{0}{0}{A\hspace*{-0.5pt}BC}$ defined in Eq.~\eqref{eq:ourrho}, we had to modify the algorithm. This is because reducing the contributions from any of the $|a_i\rangle\!\langle a_i|$ or $|\tilde{a}_i\rangle\!\langle \tilde{a}_i|$ to the original state $\rho\subtiny{0}{0}{A\hspace*{-0.5pt}BC}$ would actually \emph{increase} the purity, which would be in contradiction to the requirements of the original algorithm. The key modification is to subtract a biseparable mixture instead, which is described in detail in Appendix~\ref{appendix:2}. 
We applied the modified algorithm to each of the sampled matrices, as well as to the original reconstruction. The algorithm converged for both single-copy original density matrices, and in 95\% and 99\% of their Monte Carlo samples, respectively. In the remaining cases, we must report an inconclusive result.

Lastly, GME activation was observed experimentally. To measure the mean value of the witness, $\langle W \rangle$, we sequentially prepared all possible products $\ket{\psi}\otimes\ket{\phi}$ with $\ket{\psi},\ket{\phi}\in\bigl\{\ket{a_i}_{i=0,\ldots,7}\bigr\}\bigcup\bigl\{\,\ket{\tilde{a}_8}, \ket{\tilde{a}_9}\bigr\}$, where the first tensor factor refers to qubits $A_{1}B_{1}C_{1}$ and the second to qubits $A_{2}B_{2}C_{2}$. 

The state $\rho\subtiny{0}{0}{A_{1}\hspace*{-0.5pt}B_{1}C_{1}}\otimes\rho\subtiny{0}{0}{A_{2}B_{2}C_{2}}$ was converted into $\rho\subtiny{0}{0}{A_{1}A_{2}B_{1}B_{2}C_{1}C_{2}}$ using swap gates. For each of the $10\times 10$ constituents of that state, we then performed the Pauli measurements $M_k$ appearing in the decomposition described by Eq.~\eqref{eq:decomposedwitness} with 50 shots each. The first 16 terms $M_k$, $k=0, 1,\ldots, 15$, were measured at once by measuring all
qubits in the $Z$ basis. From the measured data, we then calculated the
witness mean and its statistical error using the vector formalism described in Appendix~\ref{appendix:3}. The estimated witness mean value is
\begin{equation}
\langle W \rangle = (-5.7 \pm 0.5)\cdot 10^{-3},
\end{equation}
which is more than eleven standard deviations below zero. This convincingly verifies the presence of GME in the two-copy state and completes our experimental demonstrations of multi-copy GME activation.\\

\noindent\emph{Discussion and Conclusion}. We have experimentally verified two-copy GME activation with state-of-the-art trapped-ion qubits. Our results thereby represent a crucial first step toward the exploration and utilization of quantum resources that are unlocked by jointly processing locally accessible subsystems of multiple copies. 
At the same time, our results highlight the challenges that will arise in attempts to harness higher levels of the GME activation hierarchy~\cite{YamasakiMorelliMiethlingerBavarescoFriisHuber2022}: One lies in the exponential growth of the number of constituents of the considered multi-copy mixed states. While our two-copy experiment required the preparation of 100 combinations, a straightforward extension to three copies would require 1000.
Another factor is the significant increase in resource requirements for witness-based GME certification and biseparability checks. 
For instance, for three-copy GME activation of the mixture given by Eq.~\eqref{eq:ourrho} with $q=0.26$, the SDP defined in Eq.~\eqref{prog:sdp} does not find any two-copy GME witness. However, a three-copy witness comprising 128 Pauli strings exists, whose evaluation requires four times more measurements compared to the two-copy witness tested here. To verify two-copy biseparability of the experimentally prepared states via the subtraction algorithm would also require demanding six-qubit quantum tomography on each copy pair.

The above obstacles are all technical in nature and can be overcome by finding simpler three-copy GME activatable states and/or streamlining the process of preparing and measuring the states used. As three-copy GME activation is a more subtle effect, the tolerance to infidelities becomes narrower. To assess its feasibility, one should also take into account the effect on the measurement uncertainty. 
Therefore, we used the available reconstruction of the two realizations of single copies and extrapolated the density matrix of the three-copy state in silico for the case $q=0.26$. The fidelity of this density matrix to the ideal theoretical state would be 0.954, leading to an expected witness value of $\langle W \rangle = (-7\pm4) \cdot 10^{-5}$ when using 50 shots per setting. Compared to the theoretical value for a perfect state of $\langle W \rangle=-8.5 \cdot 10^{-4}$, this indicates that observing three-copy GME activation is quite time-consuming, yet achievable.

Having demonstrated the feasibility of accessing GME from two copies of biseparable states, an exciting next step will be to observe and utilize GME activation on spatially separated systems in a multi-party quantum network. The states used in our demonstration are indeed typical for what one might expect in a quantum network with bipartite entanglement sources. \\


\begin{acknowledgments}
{\noindent}\emph{Acknowledgements}.
We acknowledge support from the Austrian Science Fund (FWF) through the project P 36478-N funded by the European Union - NextGenerationEU and through the EU-QUANTERA project TNiSQ (N-6001), as well as from the Austrian Federal Ministry of Education, Science and Research via the Austrian Research Promotion Agency (FFG) through the flagship project HPQC (FO999897481), the project FO999914030 (MUSIQ), and the project FO999921407 (HDcode) funded by the European Union - NextGenerationEU. O.L. acknowledges support from Palack\'{y} University, a grant no. IGA-PrF-2025-010. R.S. acknowledges the support from the Ministry of the Interior of the Czech Republic, project NU-CRYPT (VK01030193), and from the Ministry of Education, Youth and Sports of the Czech Republic, grant no. 8C22003 (QD-E-QKD) of the QuantERA II Programme that has received funding from the European Union's Horizon 2020 research and innovation programme under Grant Agreement no. 101017733. We thank Jan Provazn\'{i}k for building and maintaining the computational cluster at Palack\'{y} University, Department of Optics.\\ 

\end{acknowledgments}

{\noindent}\emph{Data availability.}
The data that support the findings of this article are openly available~\cite{datarepository}.

\newpage
\section*{Appendix: Supplemental Information}

\renewcommand{\thesubsubsection}{A.\Roman{subsection}.\arabic{subsubsection}}
\renewcommand{\thesubsection}{A.\Roman{subsection}}
\renewcommand{\thesection}{}
\setcounter{equation}{0}
\numberwithin{equation}{section}
\setcounter{figure}{0}
\renewcommand{\theequation}{A.\arabic{equation}}
\renewcommand{\thefigure}{A.\arabic{figure}}


{\noindent}In the appendix we present additional details and explicit calculations supporting our results. The appendix is structured as follows: In Sec.~\ref{appendix:1} we provide a detailed description of the two-copy GME witness. Section~\ref{appendix:2} contains additional details on the modification of the subtraction algorithm used to prove biseparability of the investigated states. Finally, Sec.~\ref{appendix:3} describes the error propagation in the witness measurement.


\subsection{Two-copy GME witness}\label{appendix:1}
\vspace*{-1.5mm}

{\noindent}In this section, we present supporting details about the witness detecting GME in two copies of state Eq.~\eqref{eq:rho}. We numerically solved the SDP in Eq.~\eqref{prog:sdp} and obtained the witness in X-matrix form. All non-zero elements of the optimal witness $W$ are given in Tab.~\ref{tab:W}. The witness can be written using GHZ basis compactly as:
\begin{align}
    W = \tfrac{1}{12}\bigg[& \sum\limits_{i\in\mathcal{K}} \left(|\mathrm{B}i\rangle\!\langle \mathrm{B}i| + |\mathrm{B}\bar{i}\rangle\!\langle \mathrm{B}\bar{i}| \right) + \sum\limits_{i = 0, 21}|\mathrm{G}i^-\rangle\!\langle \mathrm{G}i^-|\nonumber\\
    &  -\sum\limits_{i=0, 21} \left( |\mathrm{B}i\rangle\!\langle \mathrm{B}i| + |\mathrm{B}\bar{i}\rangle\!\langle \mathrm{B}\bar{i}| \right) \bigg],
    \label{W2:statedecomp}
\end{align}
where $\mathcal{K}=\{3,12,15,22,25,26\},$ the bar symbol denotes bit-wise not, and $|\mathrm{B}i\rangle=|i_5i_4\dots i_0\rangle$ is the computational basis state with $i_5 i_4 \dots i_0$ being the binary representation of $i$, and $|\mathrm{G}i^{-}\rangle = |i_5 i_4 \dots i_0 \rangle - |\bar{i}_5 \bar{i}_4 \dots \bar{i}_0 \rangle$ denotes unnormalized GHZ basis state. 


\setlength\extrarowheight{1.0pt}
\begin{table}[b]
    \caption{Non-zero elements of the fully decomposable witness $W=\sum_{ij}\mathcal{W}_{ij}|i\rangle\!\langle j|$.}
        \begin{tabular}{ *{3}{>{\centering\arraybackslash}p{2cm}} } 
        \hline\hline
            $i$ & $j$ & $\mathcal{W}_{ij}$ \\\hline
             & & \\[-8pt]
            000000 & 111111 & $-1/12$ \\
            000011 & 000011 & $\phantom{-}1/12$ \\
            001100 & 001100 & $\phantom{-}1/12$ \\
            001111 & 001111 & $\phantom{-}1/12$ \\
            110000 & 110000 & $\phantom{-}1/12$ \\
            110011 & 110011 & $\phantom{-}1/12$ \\
            111100 & 111100 & $\phantom{-}1/12$ \\
            111111 & 000000 & $-1/12$ \\
            010101 & 010101 & $\phantom{-}1/12$ \\
            011001 & 011001 & $\phantom{-}1/12$ \\
            010101 & 101010 & $-1/12$ \\
            011010 & 011010 & $\phantom{-}1/12$ \\
            100101 & 100101 & $\phantom{-}1/12$ \\
            101010 & 010101 & $-1/12$ \\
            100110 & 100110 & $\phantom{-}1/12$ \\
            101001 & 101001 & $\phantom{-}1/12$\\
            \hline\hline
        \end{tabular}
    \label{tab:W}
\end{table}

The witness operator in Eq.~\eqref{W2:statedecomp} can be further decomposed into a sum of six-fold tensor product of the Pauli operators in Eq.~\eqref{eq:decomposedwitness}. The constituent terms of the decomposition are listed in Tab.~\ref{tab:Wdecomp}.


\begin{table}[th]
    \caption{Decomposition of the witness $W$, Eq.~\eqref{W2:statedecomp}, with elements given in Tab.~\ref{tab:W} into Kronecker products of standard Pauli matrices $X, Y, Z$ and identity matrix $\openone$. The columns indicate in which basis the qubits are measured and $m_k$ gives the decomposition coefficient.}
    \begin{ruledtabular}
    \footnotesize
    \small
        \begin{tabular}{cccccccc}
            $k$ & $A_1$ & $A_2$ & $B_1$ & $B_2$ & $C_1$ & $C_2$ & $m_k$ \\ \hline
             & & & & & \\[-8pt]
            0 & $\openone$ & $\openone$ & $\openone$ & $\openone$ & $\openone$ & $\openone$ & $1$ \\
            1 &$\openone$ & $\openone$ & $\openone$ & $Z$ & $\openone$ & $Z$ & $-\frac{1}{3}$ \\
            2 &$\openone$ & $\openone$ & $Z$ & $\openone$ & $Z$ & $\openone$ & $-\frac{1}{3}$ \\
            3 & $\openone$ & $\openone$ & $Z$ & $Z$ & $Z$ & $Z$ & $1$ \\
            4 & $\openone$ & $Z$ & $\openone$ & $\openone$ & $\openone$ & $Z$ & $-\frac{1}{3}$ \\
            5 & $\openone$ & $Z$ & $\openone$ & $Z$ & $\openone$ & $\openone$ & $-\frac{1}{3}$ \\
            6 & $\openone$ & $Z$ & $Z$ & $\openone$ & $Z$ & $Z$ & $-\frac{1}{3}$ \\
            7 & $\openone$ & $Z$ & $Z$ & $Z$ & $Z$ & $\openone$ & $-\frac{1}{3}$ \\
            8 & $Z$ & $\openone$ & $\openone$ & $\openone$ & $Z$ & $\openone$ & $-\frac{1}{3}$ \\
            9 & $Z$ & $\openone$ & $\openone$ & $Z$ & $Z$ & $Z$ & $-\frac{1}{3}$ \\
            10 & $Z$ & $\openone$ & $Z$ & $\openone$ & $\openone$ & $\openone$ & $-\frac{1}{3}$ \\
            11 & $Z$ & $\openone$ & $Z$ & $Z$ & $\openone$ & $Z$ & $-\frac{1}{3}$ \\
            12 & $Z$ & $Z$ & $\openone$ & $\openone$ & $Z$ & $Z$ & $1$ \\
            13 & $Z$ & $Z$ & $\openone$ & $Z$ & $Z$ & $\openone$ & $-\frac{1}{3}$ \\
            14 & $Z$ & $Z$ & $Z$ & $\openone$ & $\openone$ & $Z$ & $-\frac{1}{3}$ \\
            15 & $Z$ & $Z$ & $Z$ & $Z$ & $\openone$ & $\openone$ & $1$ \\
            16 & $X$ & $X$ & $X$ & $X$ & $X$ & $X$ & $-\frac{1}{3}$ \\
            17  & $X$ & $X$ & $X$ & $Y$ & $X$ & $Y$ & $\frac{1}{3}$ \\
            18 & $X$ & $X$ & $Y$ & $X$ & $Y$ & $X$ & $\frac{1}{3}$ \\
            19 & $X$ & $X$ & $Y$ & $Y$ & $Y$ & $Y$ & $-\frac{1}{3}$ \\
            20 & $X$ & $Y$ & $X$ & $X$ & $X$ & $Y$ & $\frac{1}{3}$ \\
            21 & $X$ & $Y$ & $X$ & $Y$ & $X$ & $X$ & $\frac{1}{3}$ \\
            22 & $X$ & $Y$ & $Y$ & $X$ & $Y$ & $Y$ & $-\frac{1}{3}$ \\
            23 & $X$ & $Y$ & $Y$ & $Y$ & $Y$ & $X$ & $-\frac{1}{3}$ \\
            24 & $Y$ & $X$ & $X$ & $X$ & $Y$ & $X$ & $\frac{1}{3}$ \\
            25 & $Y$ & $X$ & $X$ & $Y$ & $Y$ & $Y$ & $-\frac{1}{3}$ \\
            26 & $Y$ & $X$ & $Y$ & $X$ & $X$ & $X$ & $\frac{1}{3}$ \\
            27 & $Y$ & $X$ & $Y$ & $Y$ & $X$ & $Y$ & $-\frac{1}{3}$ \\
            28 & $Y$ & $Y$ & $X$ & $X$ & $Y$ & $Y$ & $-\frac{1}{3}$ \\
            29 & $Y$ & $Y$ & $X$ & $Y$ & $Y$ & $X$ & $-\frac{1}{3}$ \\
            30 & $Y$ & $Y$ & $Y$ & $X$ & $X$ & $Y$ & $-\frac{1}{3}$ \\
            31 & $Y$ & $Y$ & $Y$ & $Y$ & $X$ & $X$ & $-\frac{1}{3}$ \\
        \end{tabular}
    \end{ruledtabular}
    \label{tab:Wdecomp}
\end{table}

To secure the fully decomposable characteristic of the witness $W$, we provide full forms of matrices $P_k$ and $Q_k$ found in the SDP. Non-zero elements of the matrices $P_{A_1A_2}$ and $P_{B_1B_2}$ are distributed only on its diagonal and are shown in Tab.~\ref{tab:Pk}. Matrix $P_{C_1C_2}$ is obtained from the previous two as $P_{C_1C_2}=|P_{A_1A_2}-P_{B_1B_2}|$. The matrices $Q_k$, $k=\{A_1 A_2,B_1 B_2,C_1 C_2\}$ contain the same diagonal elements as $P_k$ but have some extra diagonal and off-diagonal elements,
\begin{subequations}
\label{W2:Qk}
 \begin{align}
    Q_{A_1A_2}&=\,P_{A_1A_2}\\
    &\ \ \ +\tfrac{1}{12}\left(\ket{001111}-\ket{110000}\right)\left(\bra{001111}-\bra{110000}\right)\nonumber\\
    &\ \ \ +\tfrac{1}{12}\left(\ket{011010}-\ket{100101}\right)\left(\bra{011010}-\bra{100101}\right),
    \nonumber 
\end{align}
\end{subequations}\addtocounter{equation}{-1}
\begin{subequations}\addtocounter{equation}{1}\label{eq:DM matrix elements 3 qubits 2}
\begin{align}
    Q_{B_1B_2}&=\,P_{B_1B_2}\\
    &\ \ \ +\tfrac{1}{12}\left(\ket{001100}-\ket{110011}\right)\left(\bra{001100}-\bra{110011}\right)\nonumber\\
    &\ \ \ +\tfrac{1}{12}\left(\ket{011001}-\ket{100110}\right)\left(\bra{011001}-\bra{100110}\right),\nonumber \\[2mm]
    Q_{C_1C_2}&=\,P_{C_1C_2}\\
    &\ \ \ +\tfrac{1}{12}\left(\ket{000011}-\ket{111100}\right)\left(\bra{000011}-\bra{111100}\right)\nonumber\\
    &\ \ \ +\tfrac{1}{12}\left(\ket{010110}-\ket{101001}\right)\left(\bra{010110}-\bra{101001}\right).
    \nonumber
\end{align}   
\end{subequations}

\begin{table}[h]
    \caption{Non-zero elements of matrices $P_{A_1 A_2}$ and $P_{B_1 B_2}$, where the matrices are given as $P_{A_1 A_2}=\sum_i p_{A_1 A_2,i}\ketbra{i}{i}$ and $P_{B_1 B_2}=\sum_i p_{B_1 B_2,i}\ketbra{i}{i}$, respectively.}
    \begin{ruledtabular}
        \begin{tabular}{*{5}{>{\centering\arraybackslash}p{1.5cm}}}
            $i$ & $p_{A_1 A_2,i}$ &  & $i$ & $p_{B_1 B_2,i}$ \\ \hline
             & & & & \\[-8pt]
            000011 & $\frac{1}{24}$ & & 000011 & $\frac{1}{24}$ \\[1mm]
            001100 & $\frac{1}{24}$ & & 001111 & $\frac{1}{24}$ \\[1mm]
            010110 & $\frac{1}{24}$ & & 010110 & $\frac{1}{24}$ \\[1mm]
            011001 & $\frac{1}{24}$ & & 011010 & $\frac{1}{24}$ \\[1mm]
            100110 & $\frac{1}{24}$ & & 100101 & $\frac{1}{24}$ \\[1mm]
            101001 & $\frac{1}{24}$ & & 101001 & $\frac{1}{24}$ \\[1mm]
            110011 & $\frac{1}{24}$ & & 110000 & $\frac{1}{24}$ \\[1mm]
            111100 & $\frac{1}{24}$ & & 111100 & $\frac{1}{24}$ \\[1mm]
        \end{tabular}
    \end{ruledtabular}
    \label{tab:Pk}
\end{table}

\subsection{Modified algorithm for proving biseparability}\label{appendix:2}

{\noindent}This algorithm adapts the approach described in~\cite{Barreiro2010, Hofmann2014} by substituting pure separable states with biseparable mixtures. The original algorithm subtracts a pure state, separable with respect to some bipartition $k|\bar{k}$, from the original density matrix or its remainder in subsequent iterations. This pure state was chosen to overlap significantly with the remainder. Here, we introduce a key modification that leverages the knowledge of our state. Instead of finding a maximally overlapping pure state, we search for a maximally overlapping mixture of separable states. The search for a constituent separable state is facilitated by biasing the remainder towards one of the theoretical constituent states in order to find the maximally overlapping pure separable state. This is done for each of the eight constituent states $|a_{0\dots7}\rangle$ given by Eq.~\eqref{a}. These maximally overlapping states are then mixed together, optimizing the weight to achieve maximal overlap. Then we proceed as in the original algorithm.

Before we write down the algorithm, let us define a bipartition index $(k_i)$ that labels under what partition the state $\ket{a_i}$ [Eq.~\eqref{a}] is separable. 
It takes the value $A|BC$ for $i=0, 1, 2, 3$ and $AC|B$ for $i = 4, 5, 6, 7$.
The maximal number of iteration is limited with $j_{\max}$. If the algorithm does not converge with this limit, the result is inconclusive, otherwise we state that the state was biseparable.

\begin{algorithm}[H]
\caption{Biseparability certification}
\begin{algorithmic}[0]
\Require $\rho_0$, $\{ |a_i\rangle \}$, $\{k_i\}$, $j_{\max}$
\State $j \gets 0$
\State $\rho_j = \rho_0$
\While{$\mathrm{Tr}[\rho_j^2] > 1/7$ and $j<j_{\max}$}
\State $\eta \gets \mathrm{FindMixture}(\rho_j, \{ |a_i\rangle \}, \{k_i\})$
\State $\rho_{j+1} \gets \mathrm{Subtract}(\rho_j, \eta)$
\State $j \gets j +1$
\EndWhile
\If{$j < j_{\max}$}
\State \textbf{return} "$\rho_0$ is biseparable"
\Else
\State \textbf{return} "inconclusive result"
\EndIf
\Statex

\Function {FindMixture}{$\rho_{j}$, $\{ |a_i\rangle \}$, $\{k_i\}$}
\State $b \gets 10^{-3} $
\ForAll {$i = 0\dots7$}
\State $\rho'_{ji} = b|a_i\rangle\!\langle a_i| + (1-b){\rho}_j$
\State $|\psi_i^{(k_i)}\rangle \gets \arg\max\limits_{|\psi_i^{(k_i)}\rangle}
|\langle\psi_i^{(k_i)}| \rho'_{ji} |\psi_i^{(k_i)}\rangle|$
\EndFor
\State $\bm{p} \gets \arg\max\limits_{\bm{p}} \sum\limits_{i=0}^7 p_i \langle \psi_{i}^{(k_i)}|\rho_j |\psi_{i}^{(k_i)}\rangle$ \\cnstr. to $\sum p_i = 1, p_i \geq 0$
\State  $\eta \gets \sum\limits_{i=0}^7 p_i |\psi_{i}^{(k_i)}\rangle\!\langle \psi_{i}^{(k_i)}|$
\State \textbf{return} $\eta$
\EndFunction
\Statex
\Function {Subtract}{$\rho_{j}$, $\eta$}
\State $\epsilon = \arg\min\limits_{\epsilon} \frac{\mathrm{Tr}[(\rho_j - \epsilon\eta)^2]]}{\mathrm{Tr}[\hat\rho_j - \epsilon\hat\eta]}$ cnstr. to $(\hat\rho_j - \epsilon\hat\eta) \succ 0$, $\epsilon \geq 0$
\State \textbf{return} $\frac{\rho_j - \epsilon\eta}{\mathrm{Tr}[\rho_j - \epsilon\eta]}$
\EndFunction
\end{algorithmic}
\end{algorithm}

\subsection{Calculation of the witness mean and its uncertainty}\label{appendix:3}

Each of the $10\times 10$ constituents of two copies of the state defined in Eq.~\eqref{eq:ourrho} was measured using the Pauli measurements $M_k$ appearing in the decomposition
Eq.~\eqref{eq:decomposedwitness} with 50 shots each. The results are
captured by the outcome-distribution vector $\mathbf{f}_{ijk}$, where
$i$ and $j$ label the constituents of the first and second copy,
respectively. Recall that the first 16 terms $M_k$, $k=0, 1,\ldots, 15$,
were measured at once by measuring all qubits in the $Z$ basis.

Denoting the corresponding outcome-distribution vector as $\tilde{\textbf{f}}_{ij}$, the $l$-th element of the vector $\mathbf{f}_{ijk}$, $k = 0 \dots 15$, is then set equal to $(f_{ijk})_{l} = (\tilde{f}_{ij})_{l}$. The contribution of the Pauli string $M_{k}$ to the witness mean can be expressed compactly via the scalar product $\langle {M}_k \rangle_{ij} = \textbf{f}_{ijk} \cdot \textbf{h}_k,$
where $\textbf{h}_{k} = \bigotimes\limits_{\mu=1}^{6} \textbf{v}_{k\mu},$ is an auxiliary vector with $\textbf{v}_{k\mu} = (1, 1)$ if the $\mu$-th most significant symbol in the string is $\openone$, and $\textbf{v}_{k\mu} = (1, -1)$ otherwise. For the resulting witness mean we get finally
\begin{equation}
\langle W \rangle = \sum\limits_{i, j = 0}^{9}\sum\limits_{k=0}^{31}\sum\limits_{l = 0}^{63}w_{i}w_{j}m_{k}(f_{ijk})_{l}(h_{k})_{l},
\label{eq:witness_value_from_data}
\end{equation}
where $w_{i}=(1-q)/8$, $i=0,\ldots,7$, and $w_{i}=q/2$, $i=8,9$, are weights of the constituent states in the mixture in Eq.~\eqref{eq:ourrho}.

Let us derive an expression for evaluation of the statistical uncertainty of the two-copy GME witness value by propagating multinomial variances and covariances through the linear relation that forms the witness value. Due to the multi-dimensional nature of our data we use multiple indices, which we summarize in the legend in Tab.~\ref{tab:witness_formula_indices}.

The witness value is constructed as a linear combination of the measured relative frequencies. Before we continue to the derivation, let us first simplify the expression to a scalar product
\begin{equation}
\langle W \rangle = \mathbf{f} \cdot \bm{\mathcal{M}},
\end{equation}
with the relative probabilities $f_{(ijkl)}\equiv (f_{ijk})_{l}$ being the elements of a vector $\mathbf{f}$ with multi-index $(ijkl)$. Elements $\mathcal{M}_{(ijkl)}\equiv w_i w_j m_k (h_{k})_l$ describe the weights of the constituent states, as well as the witness measurements.

\begin{table}
\begin{tabular}{l p{8cm}}
\hline
\hline
\textbf{Index} & \textbf{Description} \\
$i$ & State index of copy 1 (0 to 9) \\
$j$ & State index of copy 2 (0 to 9) \\
$k$ & Witness measurement index (0 to 31) \\
$k'$ & Reduced measurement index (0 to 16) \\
$l$ & Measurement outcome index (0 to 63) \\
\hline
\hline
\end{tabular}
\label{tab:witness_formula_indices}
\caption{Indices legend.}
\end{table}

Because the first 16 witness measurements were measured at once, we introduce the vector of measured relative frequencies $\mathbf{f}'$ with elements
\begin{align}
\label{fprimed}
&f'_{(ijk'l)}\equiv\left\{\begin{array}{lll}
(\tilde{f}_{ij})_{l}& \!\textrm{for}\! & k'=0;\\
(\tilde{f}_{ijk'+15})_{l}& \!\textrm{for}\! & k'=1,\ldots,16,
\end{array}\right.
\end{align}
and a vector of weights $\bm{\mathcal{M}}'$ with elements 
\begin{align}
\label{calMprimed}
&\mathcal{M}_{(ijk'l)}'\equiv\left\{\begin{array}{lll}
w_i w_j \sum\limits_{k=0}^{15}  m_k (h_{k})_{l}& \!\textrm{for}\! & k'=0;\\
w_i w_j m_{(k'+15)} (h_{(k'+15)})_{l} & \!\textrm{for}\! & k'=1,\ldots,16.
\end{array}\right.
\end{align}

Then the dimensions of the measurement description match the dimensions of the observed data and we can write:
\begin{equation}
\langle W \rangle =\sum\limits_{i, j = 0}^{9}\sum\limits_{k'=0}^{16}\sum\limits_{l = 0}^{63}f'_{(ijk'l)}\mathcal{M}_{(ijk'l)}'\equiv\mathbf{f}' \cdot \bm{\mathcal{M}}'.
\end{equation}
We introduce the covariance matrix $\mathbb{\Sigma}$, assuming that the outcomes of a single measurement of the Pauli string $M_{k}$ obey the multinomial distribution. Then, the diagonal elements of the covariance matrix represent the variance,
\begin{equation}
\mathbb{\Sigma}_{(ijk'l),(ijk'l)} = \frac{1}{n}f'_{(ijk'l)} \left(1-f'_{(ijk'l)}\right),
\end{equation}
with $n=50$ being the number of shots per constituent state and measurement.
The off-diagonal components correspond to covariance,
\begin{eqnarray}
\mathbb{\Sigma}_{(ij k'l),(i'j'k''l')} = -\frac{1}{n}\delta_{ k'k''} \delta_{i i'}\delta_{j j'} f'_{(ijk'l)} f'_{(i'j'k''l')},\nonumber\\
\end{eqnarray}
with $\delta_{ij}$ representing the Kronecker delta.

Then, the variance of the witness mean value is
\begin{equation}
\mathrm{var}\,\langle W\rangle = \bm{\mathcal M}^{T}\left(\text{\(\mathbb{\Sigma}\)}\bm{\mathcal M}\right).
\label{eq:error_propagation_law}
\end{equation}
The covariance matrix is, however, vast (square matrix of 108 thousands rows), sparse, blocked, and impractical to store in the computer memory as a dense array. We utilize the block structure of the covariance matrix to simplify the product on the right-hand side of Eq.~\eqref{eq:error_propagation_law} by means of the formula
\begin{equation}
\text{\(\mathbb{\Sigma}\)}\,\bm{\mathcal{M}} = \operatorname{diag}\!\left(\text{\(\mathbb{\Sigma}\)}\right)\bm{\mathcal{M}} + \bm{\zeta},
\end{equation}
where the elements of the vector $\bm{\zeta}$ read as
\begin{equation}
\begin{split}
\zeta_{(ijk'l)} &= -\sum_{i'j'k''l'}
f_{(ijk'l)} f_{(i'j'k''l')} \\ & \mathcal{M}_{(i'j'k''l')} \frac{\delta_{i i'} \delta_{j j'} \delta_{k k'} (1-\delta_{l l'})}{n}.  
\end{split}    
\end{equation}
The vector $\bm{\zeta}$ can be interpreted as a term that reduces the overall variance of the results due to correlation in the multinomial distributions. 

\begin{figure}[h]
    \centering
    \includegraphics[width=0.99\linewidth]{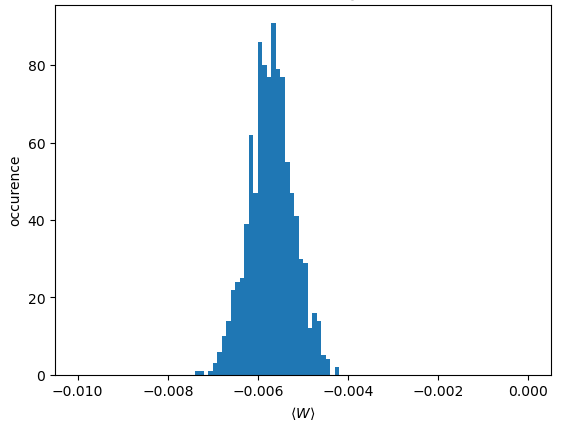}
    \caption{Histogram of two-copy witness value obtained by Monte Carlo resampling of the original tomogram.}
    \label{fig:witnesshisto}
\end{figure}

We conclude this section by stating that we used Monte Carlo resampling of the tomograms to independently evaluate the statistical uncertainty as a reference. We observed values very similar to those obtained with direct error propagation, as we can observe from the histogram in Fig.~\ref{fig:witnesshisto}.


\bibliographystyle{apsrev4-1fixed_with_article_titles_full_names_new}
\bibliography{Master_Bib_File}

\end{document}